\documentstyle[epsf,12pt,aaspp4,flushrt]{article}
\tighten

\def \ergs {ergs~s$^{-1}$~cm$^{-2}$}
\def \lognlogs {log~N~-~log~S}

\lefthead{Vikhlinin and Forman}
\righthead{CORRELATION OF X-RAY SOURCES}

\slugcomment{submitted to {\em The Astrophysical Journal Letters}}

\begin{document}

\title{Detection of the Angular Correlation of Faint X-ray Sources}

\vspace{1cm}

\author{A.~Vikhlinin\altaffilmark{1}}\affil{Space Research Institute,
Profsoyuznaya 84/32,Moscow 117810, Russia. vikhlinin@hea.iki.rssi.ru}
\and
\author{W.~Forman}
\affil{Harvard-Smithsonian Center for Astrophysics, 60 Garden street,
 Cambridge, MA 02138, USA. \\ wforman@cfa.harvard.edu}

\altaffiltext{1}{Visiting CfA}

\begin{abstract}

We have analyzed a set of deep ROSAT observations with a total sky coverage
of 40 square degrees to search for clustering of faint X-ray sources.  Using
the resulting catalog of discrete X-ray sources, we detect, for the first
time in X-rays, a positive correlation on angular scales of
0\farcm5--10\arcmin. When corrected for a bias due to limited spatial
resolution which amplifies the correlation, the observed angular correlation
function agrees well with that expected from the spatial correlation of
optically selected quasars, provided that they comprise an appreciable
fraction ($\ge 50$\%) of detected X-ray sources.

\end{abstract}


\newpage

\section{INTRODUCTION}

Models of the X-ray background composition can be strongly constrained by
the study of its arcminute-scale fluctuations. For example, Soltan \&
Hasinger (1994, hereafter SH94) have shown that a highly correlated
population of point sources cannot produce more than 35\% of the still
unresolved fraction of the X-ray background. Until now the angular
correlation of X-ray sources has not directly been measured, and the usual
approach in X-ray studies is to assume a spatial correlation model for X-ray
sources inferred from optical observations. In this {\em Letter} we report
the first detection of a positive angular correlation of X-ray sources
detected in a set of medium-deep exposure ROSAT fields covering a total area
of 40 square degrees. The correlation is detected on both small ($\lesssim
1$\arcmin) and moderate (1\arcmin--10\arcmin) angular scales.

This work continues our series of papers (Vikhlinin et al.\ 1995a,b,c,
hereafter Papers I, II, III) in which we developed an efficient method for
detection of point sources (Matched Filter, Paper I), and applied it to a
set of 130 long exposure high Galactic latitude ROSAT/PSPC images. Papers II
and III discuss the number-flux relation and average spectra of faint X-ray
sources. In the present study, we explore the spatial distribution of X-ray
sources by means of the two-point angular correlation function. We show that
the detected angular correlation is consistent with that expected if X-ray
sources are distributed in space like optically selected quasars.
We compare our results to the angular correlation of the fluctuations in
the X-ray background (e.g., Georgantopoulos et al. 1993 and SH94).

\section{DATA AND ANALYSIS}

Our sample consists of 254 medium and deep exposure ROSAT observations
all having high Galactic latitudes ($|b|>30$\arcdeg) and low
galactic absorption ($N_{\rm H} < 6\times 10^{20}$ cm$^{-2}$). When
possible, we stacked several ROSAT pointings in the same direction,
thus increasing the exposure for several fields. The data were
screened for high particle and solar scattered background periods
(Snowden et al.\ 1994).  Special precautions were taken to exclude
intervals of bad satellite aspect.  The exposure times range from $\sim
5,000$ s to $\sim 100,000$ s, with minimum detectable fluxes from
$\sim 5\times10^{-14}$ \ergs\ to $\sim 1.2\times10^{-15}$ \ergs\ for
the shortest and longest exposure fields, respectively. The effective
sky coverage is 40 square degrees at bright fluxes, begins to decrease
below $5\times10^{-14}$
\ergs, and remains significant down to at least $\sim 4\times10^{-15}$
\ergs, where the coverage is 1 square degree.

Compared to our previous study of the X-ray number-flux relation and average
source spectra (Papers II, III), we have doubled the geometrical sky
coverage, predominantly by inclusion of a large number of lower exposure
fields. On the other hand, any pointing which had a normal galaxy as its
target, which could potentially contaminate the sample by inclusion of
highly correlated and/or extended sources related to those galaxies, has
been excluded.  For the purposes of the present analysis, we used only the
central parts ($<12$\arcmin\ off-axis) of ROSAT images, excluding also a
3\arcmin\ radius circle around the targets of observations. Data analysis
was performed in exactly the same way as described in Papers I and II.

The conventional way to determine the two-point angular correlation function
(ACF) is to calculate the ratio of the number of pairs of sources in the
real data ($DD$), which fall in the separation range of
$(\theta,\theta+d\theta)$, and the number of those pairs expected when the
spatial distribution of sources is random ($RR$): $\hat{w}(\theta) =
\frac{DD}{RR}-1$.  Since the source detection efficiency decreases far
off the telescope axis (primarily because of the PSF degradation), the
appropriate way to determine $RR$ is to simulate observations with the
background and the source number-flux relation found in actual observations,
apply the detection algorithm to the simulated images, and compute $RR$ as
the number of detected source pairs found in the simulated data.

Simulations were performed according to the following procedure. The source
positions were chosen uniformly at random. Source fluxes were then
determined from the \lognlogs\ distribution measured in Paper II ($N(\;>S)
\propto S^{-1.55}$ for $S>2.2\times 10^{-14}$ \ergs\ and $N(\;>S) \propto
S^{-0.86}$ below this flux). The differential \lognlogs\ was truncated at a
limiting flux, $f_{\rm min}= 3\times10^{-18}$ \ergs, which is almost three
orders of magnitude below the detection threshold of the deepest
observation, and corresponds to a source surface density of $\sim 10$ per
PSF circle. At the limiting flux value, the number of photons per source is
much less than 1, so that the contribution from sources below $f_{\rm min}$
is indistinguishable from truly diffuse emission.  Source fluxes were
converted to counts (after accounting for the telescope vignetting function)
using the counts-to-flux conversion coefficient calculated for an absorbed
$\alpha=1$ power-law spectrum, with the absorption fixed at the appropriate
Galactic value for each ROSAT observation.  The number of source photons was
drawn from a Poisson distribution. Photons were distributed over the images
according to the PSF approximation of Hasinger et al.\ (1993).

To achieve the requirement of having the same background in the real and
simulated data, an additional diffuse component must be added to the
simulated images.  This simulates various PSPC background components --
particle, scattered solar, truly diffuse CXB, etc. The required diffuse
component was determined through an iterative procedure as follows: (0)
generate a ``diffuse map'' which is initially set to zero; (1) simulate
sources as described in the previous paragraph; (2) add the
Poisson-scattered ``diffuse map''; (3) detect point sources and
calculate the background map for the simulated image; (4) add the difference
between the background maps of the simulated and real images to the
``diffuse map''; repeat steps (1)-(4) until the difference between the
background maps of the simulated and real images is negligible.

Each ROSAT observation was simulated 5 times. The simulated data were
formatted identically to the actual observations and processed in exactly
the same manner as the real data. Several tests were used to check the
quality of the simulated data. First, the \lognlogs\ distribution determined
from simulations was found to be in good agreement with the input one.
Next, we checked that we do detect the same number of sources in the
simulated and real data (both for the whole data set and for various subsets
like short/long exposure, on-axis/off-axis etc.). In each case we found
good agreement between the real data and simulations; for example, the total
number of sources in the real data is 2158, while the number of sources
averaged over the simulations is 2199 (which is less than $1\sigma$
difference).

The ACF was calculated as $\hat{w}(\theta) = \frac{DD}{RR}-1$. Error bars
for the ACF were determined assuming Poisson statistics (Peebles 1980).  The
resulting ACF is shown in Fig. 1. In the angular separation range of
25\arcsec--100\arcsec\ there is an $\sim 4\sigma$ detection (238 pairs found
in the real data, 182 pairs are expected from simulations), and an $\sim
3.5\sigma$ detection between 100\arcsec\ and 500\arcsec\ (4513 pairs in the
real data and 4276 in simulations). The power law model $w(\theta) =
(\theta/\theta_0)^{1-\gamma} $ yields best fit parameters $\gamma = 1.7
\pm 0.3$ and a correlation angle $\theta_0 = 10^{\prime\prime} \pm
8^{\prime\prime}$ (one-parameter 68\% confidence intervals); if the value of
$\gamma$ is fixed at 1.8 (the value found for normal galaxies and optically
selected quasars), the error in $\theta_0$ is much smaller: $\theta_0 =
10^{\prime\prime} \pm 2^{\prime\prime}$.

What is the origin of the positive correlation which we detect?  Potential
contributors to the small-scale correlation include extended sources like
clusters of galaxies, nearby groups or even single galaxies of large angular
size because they can be detected as multiple point sources.  However, we
have compiled the sample so that it does not contain such objects as targets
of observations. In addition, we have searched through NED and SIMBAD and
found that very few catalogued clusters, groups or nearby galaxies are
contained in the inner field of view for the observations we used.  Visual
inspection of close pairs has shown that in only a few cases are the sources
likely to be related to a single extended source; they are not ``special''
objects such as weak ``sources'' in the wings of the PSF around very bright
sources or ghost images. Also, the detection of correlation at large
separations (100\arcsec--500\arcsec) does not support the hypothesis that
the correlation is due to the presence of a number of extended sources.
Finally, the (relatively well-constrained) power law slope of the ACF is
fully consistent with the power law slope of 1.8 found for normal galaxies
(e.g. Davis \& Peebles 1983) and the suggested value for optically selected
quasars (Shanks \& Boyle 1994).  Thus it is likely that the observed
correlation arises from clustering of QSOs, the most numerous class of faint
X-ray sources (Boyle et al.\ 1993).  Below we argue that the amplitude of
the angular correlation expected from the spatial correlation of optically
selected quasars is in good agreement with our measurements, when the latter
are appropriately corrected for the amplification bias which arises from the
limited PSPC spatial resolution.

\section{CORRECTION FOR THE AMPLIFICATION BIAS}

Unfortunately, the best fit correlation angle $\theta_0 = 10^{\prime\prime}$
obtained for the directly measured ACF of the detected sources is smaller
than the FWHM of the ROSAT PSPC PSF, which is $\sim 25$\arcsec\ on-axis.
This implies that sources separated by less than $\sim 20$\arcsec\ are
detected as a single object. Thus the distribution of detected sources is
quite different from the distribution of real sources on the sky. It is
clear that the spatial distribution of detected sources is close to the
distribution of peaks in the pattern obtained by convolution of the
distribution of real sources with the PSPC PSF. This is the origin of the
amplification bias, which results in a more clustered distribution of peaks
compared to the distribution of underlying objects.  This effect was studied
by Kaiser (1984) as applied to the correlation function of clusters of
galaxies.  Kaiser has shown that if one smoothes the galaxy distribution
with a Gaussian window of $\sim 10$ Mpc size, the correlation function of
peaks (``clusters'') is biased with respect to the correlation function of
galaxies: $\xi_{\rm clusters}(r)=A\xi_{\rm galaxies}(r)$. For reasonable
values of model parameters, the amplification factor $A$ can be as large as
$\sim 10$.  Therefore, we can expect that our results may be strongly
affected by the amplification bias.

To determine the influence of the amplification bias, we have repeated the
Monte-Carlo simulations described in section 2, with correlated input source
positions.  The sources with a power-law angular correlation function were
simulated using the 2-dimensional version of the algorithm, described in
Soneira \& Peebles (1978). We fixed the power law slope of the input
correlation function and adjusted its amplitude so that the ACF of detected
sources (measured as described in section 2) is equal to that in the real
data. For the simulations, we know the positions of both input and detected
sources, so that we can determine the influence of the amplification bias.
The ACFs of input and detected sources are shown in Fig 2. The power law
shape of the input ACF was unaffected by the PSF and the detection
algorithm, while the amplitude of the correlation function of detected
sources is increased by a factor of 2.85. To check that the difference is
indeed caused by the amplification bias, and not by an error in our
detection algorithm, we also measured the ACF of peaks in the pattern
obtained by the convolution of input source positions with the PSF.  This is
shown by open circles in Fig 2. The agreement of the ACF for detected
sources and peaks is excellent. The success of this test makes us confident
that we are indeed dealing with the amplification bias.  The correction
factor of 2.85 must be applied to the measured ACF in order to obtain the
unbiased estimate.  With this correction, the correlation angle for the
underlying source population is $\theta_0 = 4^{\prime\prime}$. In the
following section we compare this corrected ACF with that expected for the
sources clustered like optically selected quasars.

\section{COMPARISON WITH OPTICAL AND OTHER X-RAY STUDIES}

Most studies of QSO clustering show that quasars are indeed strongly
clustered at $z\sim1-2$. Although the reports on clustering evolution are
controversial, there is a general agreement about the clustering amplitude
at these redshifts -- at a comoving scale of $10 h^{-1}$ Mpc the amplitude
of the two-point correlation function is $\sim 1$ (see Shanks \& Boyle 1994
for a recent analysis). In the calculations below, we make a common
assumption that the spatial two-point correlation function at any redshift
is given by
\begin{eqnarray} \label{eq-spcorr}
\xi(r,z) & = & (r/r_0)^{-\gamma}\; (1+z)^{-3-\varepsilon}
\end{eqnarray}
with the parameter $\varepsilon$ describing clustering evolution, for
example $\varepsilon\simeq 3-\gamma \simeq -1.2$ corresponds to the
clustering which is constant in comoving coordinates. For a power law
spatial correlation function, the angular correlation function is also a
power law $w(\theta) = (\theta/\theta_0)^{1-\gamma}$ with amplitude (Peebles
1980)
\begin{eqnarray}\label{eq-sp2a}
\theta_0^{1-\gamma} & = & r_0^{-\gamma}\; H_\gamma \;
\left(\frac{H_0}{c}\right)^\gamma \;
\frac
{\int_0^\infty
y^{5-\gamma}\,dy\,\phi(y)^2\,(1+z(y))^{-3-\varepsilon+\gamma}/F(y)}
{[\int_0^\infty y^2\,dy\,\phi(y)/F(y) ]^2},
\end{eqnarray}
where $H_\gamma=\Gamma(\case{1}{2})\,\Gamma(\case{\gamma-1}{2})/\,
\Gamma(\case{\gamma}{2})$, and $F(y)=[1-\left(\frac{H_0 a_0
x}{c}\right)^2(\Omega-1)]^{1/2}$. The parameter $y$ is related to the
coordinate  distance and redshift through
\begin{eqnarray} \label{eq-y2z}
y & = & H_0 a_0\, \frac{x}{c} \;\; = \;\;
2 \;\frac
{(\Omega-2)(1+\Omega z)^{1/2}+2-\Omega+\Omega z}
{\Omega^2(1+z)},
\end{eqnarray}
and $\phi(y)$ determines the fraction of sources observable at the given $y$
(or $z$), i.e.\ those with observed fluxes greater than the detection
threshold $S_{\min}$. The cosmological model enters through the $z$ -- $y$
relation (eq. \ref{eq-y2z}), volume factor $F(y)$, and the selection
function $\phi(y)$. The latter also depends on the source luminosity
function evolution. Boyle et al.\ (1993) derive the cosmological evolution
of QSOs in the form of pure luminosity evolution, with the characteristic
luminosity scaling with redshift as $(1+z)^\beta$. For an object with
luminosity $L(E)$ at redshift $z$, the observed energy flux is
$f=L(E(1+z))(1+z)/4\pi D_L^2$, where $D_L$ is the luminosity distance. For
the power law energy spectrum $E^{-\alpha}$ with $\alpha=1$ (which is close
to what is observed, see Paper III), $L(E(1+z))(1+z)=L(E)$ so that the
minimum luminosity is related to the minimum detectable flux simply as
$L_{\rm min} = 4\pi\, D_L^2 \, f_{\rm min}$. The selection function can be
written as
\begin{eqnarray}
\phi(y) & = & \int_{L_{\rm min}}^\infty \Phi(L,z(y))\, dL \;\; = \;\;
\int_{4\pi\, D_L^2 f_{\rm min}/R(z)}^\infty \Phi(L,0)\, dL,
\end{eqnarray}
where $\Phi(L,z)$ is the QSO luminosity function at redshift $z$, and $R(z)$
is the ratio of the typical luminosity at a redshift $z$ to that at the
present epoch. Given the spatial correlation function (eq. \ref{eq-spcorr}),
the present day QSO luminosity function $\Phi(L,0)$ and the evolution rate
$R(z)$, eqns.\ 2-4 enable one to calculate the angular correlation
amplitude, $\theta_0$. Of numerous parameters in eqns.\ 1-4, the angular
correlation function most strongly depends on the spatial clustering scale
$r_0$, the rate of clustering evolution $\varepsilon$, and the power law
slope $\gamma$. There is an indication that the value of $\gamma$ is almost
``universal'', i.e.\ clustering of different classes of objects like
galaxies, galaxy groups, clusters of galaxies, QSOs, is well described by a
power law with indices close to 1.8, while the clustering scales $r_0$ are
very different (see Bahcall 1988 for a review).  As we showed aboved,
$\gamma=1.8$ is also in good agreement with the angular correlation of X-ray
sources we report in this {\em Letter}. We hence fix $\gamma$ at the value
of 1.8 and calculate the angular correlation function for different values
of $r_0$ and $\varepsilon$.  We then compare the predicted and measured ACF,
to constrain the parameters of the spatial correlation function.  We assume
that $\Omega=1$ and the luminosity function and its evolution are those
given by Boyle et al.\ (1993). The 90\% confidence regions of parameters,
calculated for two values of the QSO fraction,\footnote{
\normalsize If the fraction of
clustered sources $f_0 < 1$, and other sources are randomly distributed, the
ACF of the total sample equals the ACF of clustered sources times $f_0^2$;
if there are two populations both clustered, but independent from each
other, the ACF must be multiplied by a factor $f^2$, $f_0<f<1$. In our
calculations we assume that QSOs are clustered, while other sources are not.}
are shown in Fig. 3. For simplicity of comparison of our
results with the analysis of SH94, the clustering scale in Fig. 3 is expressed
in units of the correlation length at $z=1.5$. The grey-shaded region
in this figure shows the range of clustering parameter values measured at
$z=1-1.5$ for optically selected quasars (Shanks \& Boyle 1994).  For any
reasonable fraction of quasars in our X-ray sample, $0.5 \le f \le 1.0$, the
X-ray and optical data are in good agreement.

We also compare our results with other studies of the X-ray background
fluctuations. The angular correlation function of the {\em unresolved}
background in ROSAT Deep Survey fields has been measured by Georgantopoulos
et al. (1993) and SH94 who detect a positive correlation on small angular
scales ($<3$\arcmin). At larger separations rather tight upper limits on the
angular correlation are obtained, and the general consensus (e.g., Carrera
\& Barcons 1992, Danese et al. 1993, SH94) is that a population of sources
which are clustered like QSOs cannot produce more than $30\%-50\%$ of the
{\em still unresolved} XRB. For example, Fig. 8 of SH94 shows that the
isotropy of the residual XRB requires that a population of sources which
contribute 100\% of the residual XRB must have a correlation length $x_0 \le
2 h^{-1}$ Mpc at $z=1.5$. This is significantly below the correlation length
of both QSOs and the X-ray sources in our sample.  Thus, there is a
considerable discrepancy: on the one hand, directly detected X-ray sources
are highly correlated (their clustering is consistent with $\sim 80\%$ of
them being spatially distributed like QSOs), while the residual XRB (after
subtraction of detected sources) appears very smooth. This may imply
significant changes in the nature of sources contributing to the XRB at
faint flux levels (close to the sensitivity limit of ROSAT Deep surveys).
Narrow emission line galaxies identified in Deep surveys (e,g., Boyle et al.
1995) may be promising candidates for a new population. Also, one cannot
exclude a significant contribution of truly diffuse emission to the
unresolved XRB.  Hasinger et al. (1993b) showed that the contribution of a
diffuse component up to 25\% of the total XRB flux is not excluded. Wang \&
McCray (1993) found that 40\% of the XRB in the $0.5-0.9$ keV band arises
from a diffuse thermal component.  Both results indicate that a significant
fraction of the unresolved XRB may be truly diffuse, thus making it very
smooth compared to the clustering found for detected sources.

We have made a rough estimate of the contribution of the observed
point source correlation to that reported by SH94 for the unresolved
XRB since we detect sources below their threshold (see Paper II).
SH94 gave a $2\sigma$ limit of 30-35\% for the fraction of unresolved
sources contributing to the XRB that could be clustered like quasars
based on analysis of images with a typical flux limit of about
$1\times10^{-14}$ \ergs.  Above this threshold $\sim 40$\% of the soft
XRB is resolved into point sources. The sources in the flux range
$1.5-10 \times10^{-15}$ \ergs\ (i.e. between the typical sensitivities
of the deepest field and SH94) comprise $\sim30$\% of the XRB.  If
these faint sources are clustered not significantly less than the
majority of our sources (around a flux of $10^{-14}$ \ergs), a large
fraction of the maximum correlation allowed by SH94's limit for the
``residual'' XRB must be accounted for by the sources we detect.  If
bright and faint sources exhibit the same clustering, SH94's $2\sigma$
upper limit for the correlation in the ``unresolved'' XRB is produced.
In this case the XRB below $1.5\times10^{-15}$ \ergs, which comprises
$\sim35$\% of the total XRB, should exhibit essentially no clustering
since we have already observed the quasar-like, clustered population
allowed by the SH94 limit.

\section{CONCLUSIONS}

We report on the first detection of a positive angular correlation derived
from the distribution of individual, faint X-ray sources. This correlation
is detected with high statistical significance on both small ($\lesssim
1$\arcmin) and moderate (1\arcmin--10\arcmin) angular scales. The two-point
angular correlation function is well described by a power law
$(\theta/\theta_0)^{1-\gamma}$, with $\gamma=1.7\pm0.3$ which is remarkably
close to the corresponding value for normal galaxies and optically selected
quasars. We show that the amplitude of correlation is also consistent with
the hypothesis that X-ray sources are spatially distributed in the same way
as optically selected quasars.

\acknowledgements

We thank S. Murray for his comments and continued interest and support
of this project and we thank C. Jones for comments on the manuscript.
We thank the referee, K. Jahoda, for helpful comments and a critical
reading of the manuscript. This work was supported by NASA contract
NAS8-39073. We acknowledge support from SUN Microsystems through a
hardware grant.  A.V. thanks CfA for its hospitality during his visit.
A.V. was supported by the RBRF grant 95-02-05933.

\newpage

\section*{Figure Captions}

\vspace{.5cm}

Fig.~1.---Angular two-point correlation function of X-ray sources. The
best fit power law model $(\theta/\theta_0)^{1-\gamma}$ is shown by
the dashed line.  The power law slope is relatively well constrained
and $\gamma$ is remarkably close to that for normal galaxies and
optically selected quasars (1.8).

\vspace{.5cm}

Fig.~2.---Comparison of the simulated and measured ACF. The source positions
were simulated so that the ACF of detected sources in simulations (open
circles) is equal to that of sources detected in the real data (fit is shown
by the solid line).  The ACF of input positions is shown by small solid
squares; the dashed line is the power law $\theta^{-0.8}$ fit to the input
ACF. The shapes of the input and measured ACF are the same, while the latter
has a factor of 3 higher normalization. To verify that the difference is due
to the amplification bias, the input source positions were convolved with
the PSPC PSF, and the ACF of peaks in the convolved pattern has been
measured (large solid circles).
\vspace{.5cm}

Fig.~3.---The 90\% confidence intervals for the correlation length $x_0$ at
$z=1.5$ (in comoving coordinates) and the evolution parameter $\varepsilon$.
Assuming 1) a spatial correlation function of the form $\xi(r,z) =
(r/r_0)^{-\gamma}\; (1+z)^{-3-\varepsilon}$, 2) a source luminosity function
and evolution that of X-ray selected quasars as given by Boyle et al.\
(1993), and 3) $\Omega=1$, the angular correlation function was calculated
for two values of the clustered population fraction $f= 0.6, 1.0$ and
compared with the measured ACF (Fig. 1) by means of a $\chi^2$-test. The
grey-shaded region shows the range of values of $x_0$ measured for optically
selected quasars (Shanks \& Boyle 1994). X-ray and optical data are in good
agreement if the QSO fraction is 60\%-80\%, close to what is observed.


\newpage
\pagestyle{empty}

{\bf Fig. 1}
{\vspace{3cm}
\begin{figure}[h]
\plotone{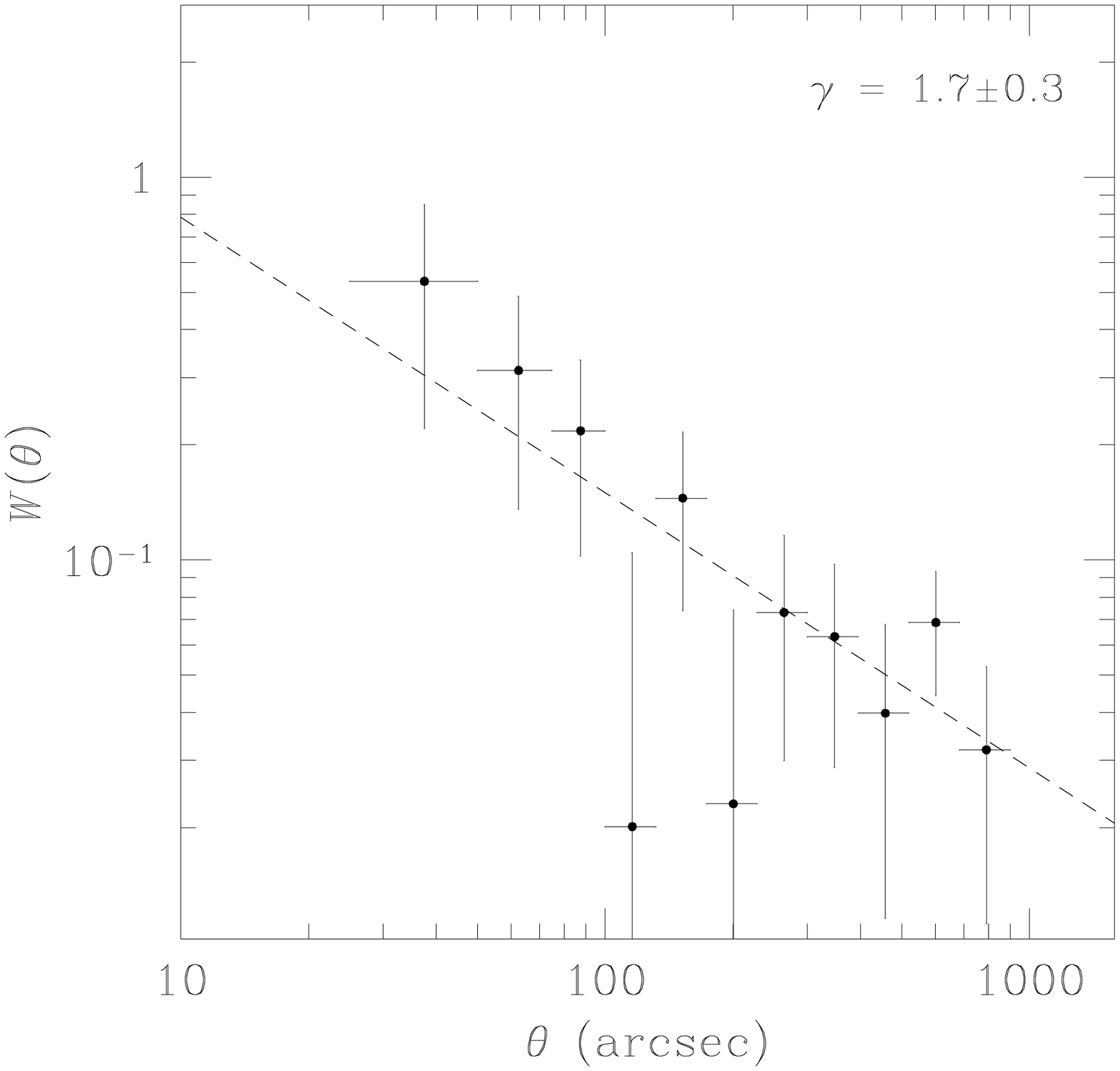}
\end{figure}

\newpage
\pagestyle{empty}

{\bf Fig. 2}
\vspace{3cm}
\begin{figure}[h]
\plotone{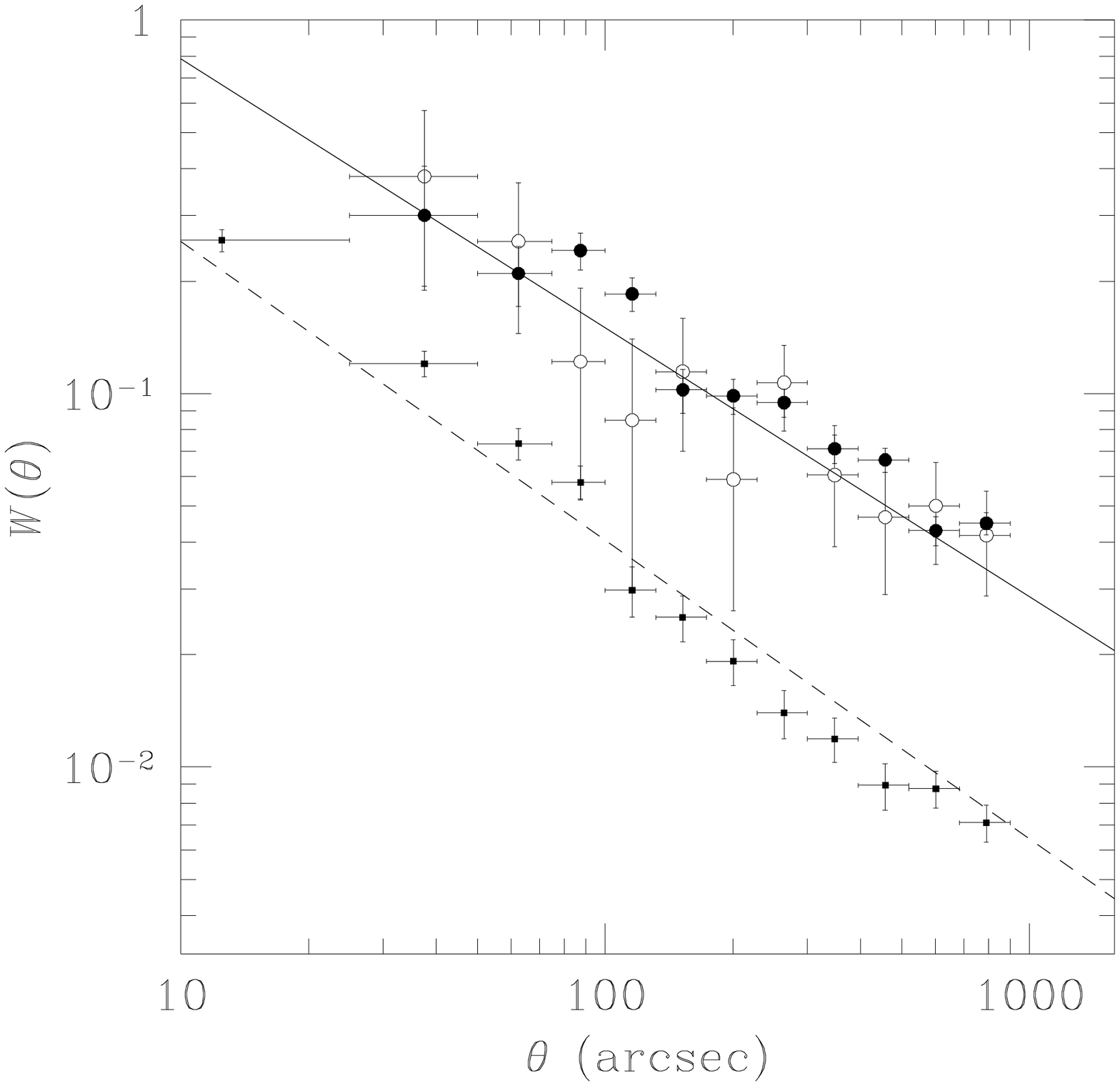}
\end{figure}

\newpage
\pagestyle{empty}

{\bf Fig. 3}
\vspace{3cm}
\begin{figure}[h]
\plotone{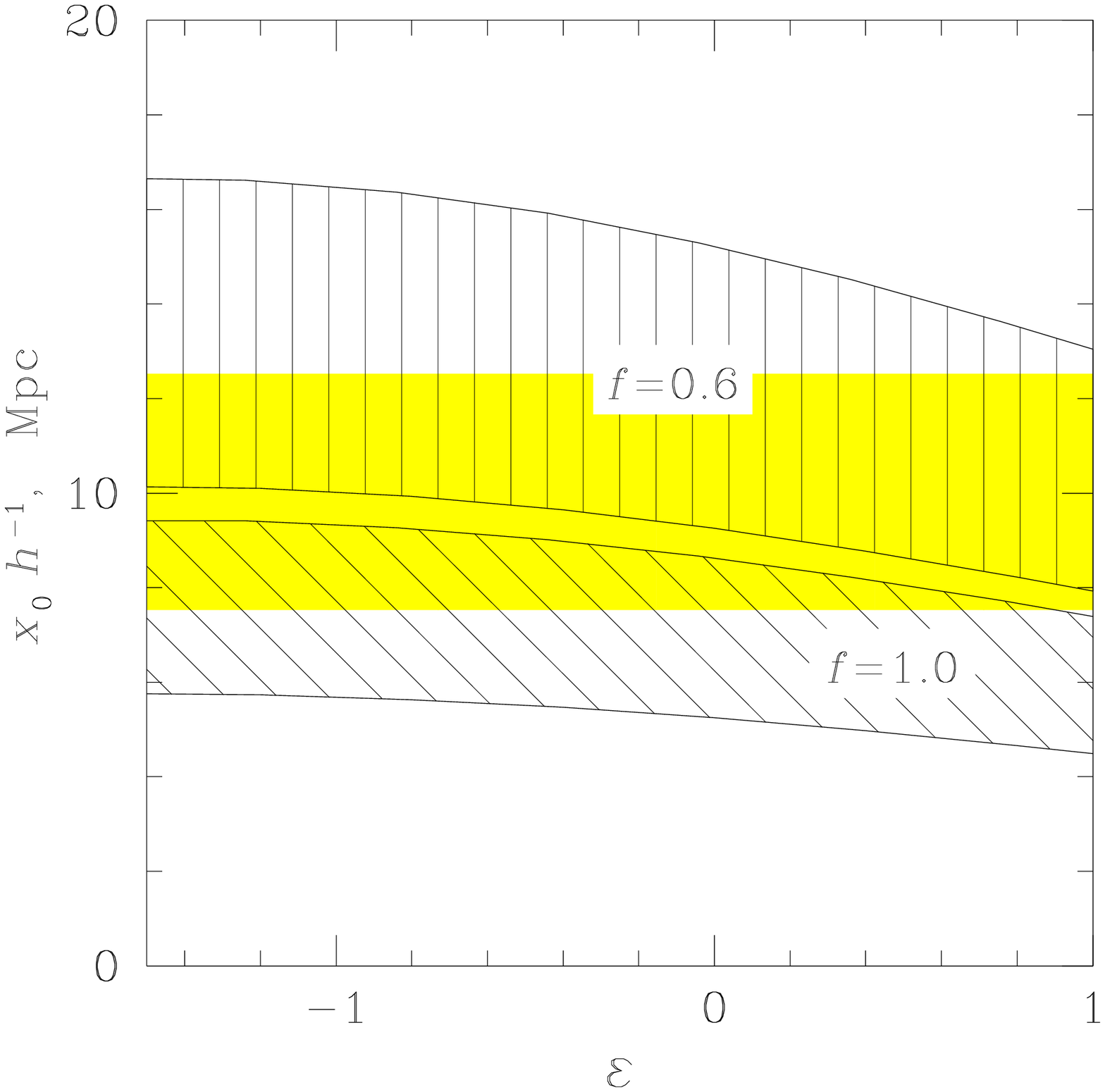}
\end{figure}

\end{document}